\documentstyle[12pt,aaspp4,flushrt]{article}
\newcommand{\lsim}{\raisebox{0.3mm}{\em $\, <$} \hspace{-3.3mm}
\raisebox{-1.8mm}{\em $\sim \,$}}

\begin{document}

\vspace*{0.5cm}

\title{A Criterion for Photoionization of Pregalactic Clouds
Exposed to Diffuse Ultraviolet Background Radiation}

%{\Large Effects of Photoionization on the Formation of Dwarf Galaxies} \\

\author{Yukiko Tajiri\altaffilmark{1}
and Masayuki Umemura\altaffilmark{2}}

\received{December 4, 1997}
\accepted{February 11, 1998}

\begin{center}
%{\it Received December 4, 1997; accepted February 11, 1998}\\[1cm]
To appear in {\it The Astrophysical Journal}
\end{center}

\altaffiltext{1}{Institute of Physics, Tsukuba University, Tsukuba,
Ibaraki 305, Japan; tajiri@rccp.tsukuba.ac.jp}

\altaffiltext{2}{Center for Computational Physics, Tsukuba University, Tsukuba,
Ibaraki 305, Japan; umemura@rccp.tsukuba.ac.jp}

%\newpage
\begin{abstract}
To elucidate the permeation of cosmic ultraviolet (UV) background radiation
into a pregalactic cloud and subsequent ionization, 
the frequency-dependent radiative transfer equation is solved, coupled 
with the ionization process, for a spherical top-hat cloud
composed of pure hydrogen. 
The calculations properly involve scattering processes of ionizing photons
which originate from radiative recombination.
As a result, it is shown that the self-shielding, although it
is often disregarded in cosmological hydrodynamic simulations,
could start to emerge shortly after the maximum expansion 
stages of density fluctuations. 
Quantitatively, the self-shielding is prominent
above a critical number density of hydrogen which is given by 
$n_{crit}=1.4\times 10^{-2}{\rm cm}^{-3} 
\left( M/ 10^8 M_\odot \right)^{-1/5}
I_{21}^{3/5}$
for $10^4$K gas,
where $M$ is the cloud mass and the UV background intensity 
is assumed to be 
$I_\nu = 10^{-21} I_{21} (\nu/ \nu_L)^{-1}  
         {\rm erg \, cm^{-2} \, s^{-1} \, ster^{-1} \, Hz^{-1}}$
with $\nu_L$ being the Lyman limit frequency. 
%$n_{crit}$ is higher for higher temperature.
The weak dependence of $n_{crit}$ upon the mass is worth noting.
The corresponding critical optical depth ($\tau_{crit}$)  
turns out to be independent of either $M$ or $I_{21}$,
which is $\tau_{crit}=2.4$ for $10^4$K gas.
The present analysis reveals that 
the Str\"{o}mgren approximation leads to overestimation of the
photoionization effects.
Also, the self-shielded neutral core is no longer sharply separated from
surrounding ionized regions;
a low but noticeable degree of ionization is caused 
by high energy photons even in the self-shielded core. 
The present results may be substantial
on considering the biasing by photoionization against low-mass galaxy formation.
\end{abstract}

%\newpage

\section{Introduction}

Recently, a recalcitrant problem on galaxy formation has been pointed out
that low-mass galaxies are overproduced as compared with observations 
in the context of the hierarchical bottom-up theory of galaxy formation
(e.g. White \& Frenk 1991; Kauffman, White, \& Guideroni 1993; Cole et al. 1994). 
Hence, some process that inhibits the formation of low-mass galaxies is required. 
Photoionization has been considered as one of such mechanisms
(Dekel \& Rees 1987; Babul \& Rees 1992; Efstathiou 1992; 
Chiba \& Nath 1994; Thoul \& Weinberg 1996; Quinn, Katz, \& Efstathiou 1996). 
In photoionized media, the cooling efficiency is dramatically
reduced in the temperature range of 
$10^4 {\rm K} < T < 10^5 {\rm K} $.
Also, a bulk of energy could be carried into a cloud, so that 
the enhanced thermal pressure could suppress the gravitational collapse of
a subgalactic cloud with the virial temperature lower than
several $10^4 {\rm K}$, i.e., $M\lsim 10^9 M_\odot$ in gas mass
(Umemura \& Ikeuchi 1984, 1985; Ikeuchi 1986; Rees 1986; 
Bond, Szalay, \& Silk 1988; Steinmetz 1995; Thoul \& Weinberg 1996).
In cosmological hydrodynamic simulations, 
an optically thin medium against ionizing photons has been 
mostly assumed so far 
(Umemura \& Ikeuchi 1984, 1985; Thoul \& Weinberg 1996; 
Quinn, Katz, \& Efstathiou 1996).
In semi-analytic approaches, naive analytic corrections  
for opacity effects have been made 
based on the optical depth criterion (Efstathiou 1992)
or the Str\"{o}mgren approximation (Chiba \& Nath 1994).
In the case of interstellar clouds, it is claimed that 
the Str\"{o}mgren approximation could be misleading against the real effect of
the penetration of diffuse UV
(Flannery, Roberge, \& Rybicki 1980; Maloney 1993), and
therefore the radiative transfer equation should be solved properly.
However, as far as we know, no study hitherto has been made on determining
the ionization structure inside a pregalactic cloud 
by solving radiative transfer equation for diffuse UV photons.
Hence, the effects of photoionization on the evolution of a pregalactic
cloud have not been assessed satisfactorily.
In this paper, we solve the radiative transfer of 
diffuse UV radiation coupled with ionization process
to elucidate the self-shielding 
of pregalactic clouds from UV background radiation
and provide a practical criterion for the self-shielding.

\section{Radiative Transfer with Ionization Process}

%\subsection{Ionization Process}

We assume for simplicity a spherical top-hat (uniform) density
distribution of a cloud, and place 100 radial meshes for solving radiation
transfer.
Also, the cloud is assumed to be composed of pure hydrogen
so that we could readily compare the results of
the frequency-dependent radiative transfer with an analytic estimate,
although we should keep in mind that helium of cosmic abundance
could alter the ionization degree maximally by order 10\%
(Osterbrock 1989; Nakamoto et al. 1997). 

The intensity of UV background radiation at high redshifts is
inferred from so-called proximity effect of Lyman $\alpha$ absorption 
lines in QSO spectra (Bajtlik, Duncan, \& Ostriker 1988; Giallongo et al. 1996).
The observations require
the diffuse UV radiation to be at a level of $I_{\nu_L,0}=10^{-21\pm 0.5}$
ergs cm$^{-2}$ s$^{-1}$ ster$^{-1}$ Hz$^{-1}$  at the hydrogen Lyman edge
at $z=1.7-4.1$. 
In this paper, we assume the specific intensity of UV background as
$I_{\nu,0} = 10^{-21} I_{21} (\nu/\nu_L)^n$ 
ergs cm$^{-2}$ s$^{-1}$ ster$^{-1}$ Hz$^{-1}$, where
we set $n=-1$ and vary $I_{21}$ in the range of $0.1 < I_{21} < 2$.
As for ionization process,
we presume the ionization balance, because the ionization 
or recombination timescale for 
clouds with density of interest is much shorter than the dynamical timescale.
The equation of the ionization balance is
\begin{equation}
\Gamma^{\gamma} {\chi_{HI}}+\Gamma^{ci} {\chi_{HI}} (1-\chi_{HI} ) n
=  \alpha_A(T)   (1-\chi_{HI} )^2 n,
\end{equation}
where $\chi_{HI}$ is the fraction of neutral hydrogen,
$\Gamma^\gamma$ is the photoionization rate,
$\Gamma^{ci}$ is the collisional ionization rate, 
$n$ is the hydrogen number density, and
$\alpha_A(T)$ is the total recombination coefficient 
to all bound levels of hydrogen.
$\Gamma^\gamma$ is given by
\begin{equation}
\Gamma^{\gamma} = \int^\infty _{\nu_L }d\nu \int^{4\pi} _0 d\Omega
               \frac{I_\nu (r)}{h\nu } a_{\nu }, \label{photoion}
\end{equation}
with the photoionization cross section being
$a_\nu =6.3 \times 10^{-18} (\nu_L/\nu)^3[{\rm cm^2 }]$, where
the local UV intensity $I_\nu (r)$ at radius $r$
is determined by solving transfer equation. 
$\Gamma^{ci}$ is given by
$
\Gamma^{ci} = 1.2 \times 10^{-8}T_4^{1/2} e^{-15.8/T_4}[{\rm cm^3 s^{-1}}]
$
with $T_4 \equiv T/10^4{\rm K}$. We assume $T_4=1$ 
except where other values are specified.
$\alpha_A(T)$ is well fit by
$
\alpha_A(T) = 2.1 \times 10^{-13}  T^{-1/2}_4 \phi(16/T_4) 
[{\rm  cm^3 s^{-1}} ],
$
where $\phi(y)= 0.5(1.7+\ln y +1/6y ) {\rm ~for~} y\geq 0.5$ or 
$y(-0.3+1.2 \ln y )+y^2 (0.5-\ln y) {\rm ~for~} y<0.5$
(Sherman 1979).

%\subsection{Radiative Transfer}

Photoionization and recombination processes can be respectively regarded 
as extinction and emission with respect to ionizing photons. 
Thus, the radiative transfer equation for ionizing photons 
is described as
\begin{equation}
\frac{dI_\nu }{ds} = -\chi_\nu I_\nu + \eta_\nu,
\end{equation}
where $\chi_\nu$ is the extinction coefficient 
($ \chi_\nu = a_\nu n \chi_{HI}$) and $\eta_\nu$ is the emissivity.
If a free electron recombines directly to the ground state of hydrogen, 
the emitted photon has enough energy 
to cause further photoionization. 
But, when an electron is captured to an excited state
of hydrogen, the emitted photon does not have enough energy to ionize hydrogen, 
because the kinetic energy of a free electron 
is of order 0.1 Ryd for $\sim 10^4$ K gas.
Thus, the former process is regarded as scatterings which provides 
the emissivity, while the latter process is pure absorption.
The effective scattering albedo is given by
$\omega=[\alpha_A(T)-\alpha_B(T)]/\alpha_A(T)$, which is 
0.4 at $10^4$ K (e.g. Osterbrock 1989), where 
$\alpha_B(T)$ is the recombination coefficient to all excited levels 
of hydrogen.
Hence, we set $\eta_{\nu} = 0~{\rm for~} \nu > \nu_L$, and
$\eta_{ \nu_L } = h \nu \omega \alpha_A n_e n_p/4 \pi \delta \nu 
~{\rm for~}\nu = \nu_L $ 
where $\delta \nu =kT/h$ and $n_p$ is the proton number density.

Taking account of frequency-dependence of the emergent UV intensity,
it is convenient to divide the photoionization rate into two parts:
$
\Gamma^\gamma = \Gamma^\gamma _{\nu_L }  + \Gamma^\gamma_{\nu>\nu_L}.
$
As for Lyman limit photons, the transfer equation
is the integro-differential equation which includes a source term
by scattering processes of recombination photons.
Thus, the equation is numerically solved with including
iterative procedure.
We, without invoking on-the-spot approximation which could be misleading
for large mean-free-path photons, solve the equation
by means of an impact parameter method of high accuracy, which
allows us to treat diffuse photons correctly (Stone, Mihalas, \& Norman 1992).
We deal with 156 impact parameters for light rays.
Also, in order to converge the intensity, we employ  
the Lambda-iteration method (e.g. Mihalas \& Mihalas 1984).

In the range of $\nu>\nu_L$,
since photoionization is regarded as pure absorption,
the UV intensity is obtained just by $I_\nu = I_{\nu,0 } e^{-\tau_\nu }$,
where $I_{\nu,0}$ is the boundary intensity 
and $\tau_\nu$ is the ionization optical depth.
So, $\Gamma^\gamma_{\nu>\nu_L}$ is figured out by
\begin{equation}
\Gamma^\gamma_{\nu>\nu_L} = \int_0^{4 \pi} 
d\Omega \int^\infty_{\nu_L + \delta \nu } 
      d\nu \frac{I_{\nu,0} e^{-\tau_\nu }}{h\nu} a_{\nu_L} 
                          \left(\frac{\nu_L }{\nu}\right)^3 .
\end{equation}
Considering that the optical depth is $\tau_\nu = \tau_{\nu_L} (\nu_L/\nu )^3
n\chi_{HI}$ with the Lyman limit optical depth $\tau_{\nu_L}$
and that the assumed boundary intensity is $I_{\nu,0} = I_{\nu_L, 0} (\nu_L/\nu)$,
the above equation can be analytically integrated 
using the incomplete gamma function $\gamma$ if $\chi_{HI}$ is given: 
\begin{equation}
\Gamma^\gamma_{\nu>\nu_L} = 
\int_0^{4 \pi} d\Omega \frac{ a_{\nu_L} I_{\nu_L ,0 } }{ h }
        \cdot \frac{\gamma (4/3 ,\, \tau_{\nu_L})}{3 \tau_{\nu_L}^{4/3} } 
	. \label{hardphotoion}
\end{equation}
With use of this integration, the overall procedure to solve the transfer 
equation is as follows:
\begin{enumerate}
\item  
Initially, give the cloud mass $M$ and the radius $R$ (therefore the density),
and set $\chi_{HI}$ by assuming optically thin medium. 
Specify the boundary UV intensity by $I_{21}$.
\item 
Solve the transfer equation at $\nu=\nu_L$ for given $\chi_{HI}$, 
and calculate $\Gamma^\gamma _{\nu_L }$. 
Obtain $\Gamma^\gamma _{\nu>\nu_L }$ analytically
by equation (\ref{hardphotoion}).
Then, get the total photoionization rate $\Gamma^\gamma$.
\item
Solve ionization equilibrium using above-obtained $\Gamma^\gamma$, and
thereby renew $\chi_{HI}$.
\item 
Continue steps 2 and 3 until $\chi_{HI}$ converges at a level 
of relative error of $10^{-6}$. (Typically 100 iterations are performed.)
\end{enumerate}

Note that the analytical integration (\ref{hardphotoion}) 
with respect to frequencies enables us to
reduce the computational cost dramatically on solving such 
a frequency-dependent transfer equation including scatterings.
The validity of this method is confirmed by exactly solving 
the transfer equation
with using a number of meshes for frequencies.

\section{Numerical Results}

We consider the cloud mass range of $M=10^{5-9} M_\odot $ and vary the radius 
in the range of $R= 0.1 - 25 {\rm kpc}$.
Fig.1  shows the growth of self-shielded regions when a cloud of 
$10^8M_\odot$ is contracting, embedded in
the UV background of $I_{21}=1$.
The self-shielding is prominent when $R \lsim 4$kpc.
It is noted that 
the distributions of the HI fraction are a gradual function of radii
even in the self-shielded stage, where
we cannot recognize a clear boundary 
between the neutral core and the ionized envelop, and also a low but
noticeable ionization is left in the self-shielded regions. 
Such distributions seem to be realized by the high frequency photons 
far above the Lyman edge which could 
permeate into deeper regions due to the smaller cross section for ionization. 
To ensure this conjecture, 
we solve the transfer solely for the Lyman edge photons 
to obtain $I_{\nu_L}(r)$ and 
tentatively set the form of UV radiation spectrum to be
$I_\nu(r) =I_{\nu_L}(r) (\nu_L/ \nu )$ at any radius
(which implies that photons of higher frequencies are absorbed with
the same ionization cross section as that at the Lyman edge).
We can see an outstanding difference between two cases as shown in Figure 1.
In the tentative case, we can see a steep inward increase of
neutral fraction and a very sharp transition from ionized regions 
to neutral regions.

Figure 2 shows the neutral hydrogen fraction at the center as
a function of cloud size.
We see that the central $\chi_{HI}$ varies abruptly at a certain 
critical size. Here, we define the critical radius $(R_{crit})$ 
at which the HI fraction at the center drops just below 0.1.
In Figure 3, we plot the critical radii obtained from all the 
numerical results as a function of UV background intensity. 
Different symbols represent different mass of gas clouds.
All the results can be remarkably well fitted by a simple formula 
which is a function of the cloud mass and the UV background intensity; 
\begin{equation}
R_{crit}=4.10{\rm kpc} \left(\frac{ M }{ 10^8 M_\odot } \right)^{2/5} 
I_{21}^{-1/5}. \label{rcrit}
\end{equation}
Equivalently, the corresponding critical number density of the cloud is
\begin{equation}
n_{crit}=1.40\times 10^{-2}{\rm cm}^{-3} 
\left(\frac{ M }{ 10^8 M_\odot } \right)^{-1/5}
I_{21}^{3/5}.
\end{equation}
It is worth noting that the critical density is quite weakly dependent upon $M$.
If the cloud is highly ionized and optically thin, the neutral fraction should be 
$\chi_{HI,0}=0.15nI_{21}^{-1}$ with neglecting collisional ionization.
Then, we define the critical optical depth $\tau_{crit}$ at the Lyman edge
as a measure such that the self-shielding becomes effective:
$\tau_{crit}\equiv n_{crit}\chi_{HI,0}a_{\nu_L}R_{crit}=2.4$,
which turns out to be independent of not only $M$ but also $I_{21}$.
Hence, such a simple criterion of $\tau=1$ as adopted by Efstathiou (1992)
is found heuristically to be practical in order to assess 
the self-shielding for $\sim 10^4$ K clouds. 

For clouds of different temperature, we have found that
$R_{crit}$ is scaled by $\alpha_B(T)^{1/5}$, and therefore 
$n_{crit} \propto \alpha_B(T)^{-3/5}$.
Resultingly, $\tau_{crit}$ is scaled by 
$\alpha_A(T)/\alpha_B(T)[=(1-\omega)^{-1}]$.
Both $\alpha_A(T)$ and $\alpha_B(T)$ are decreasing functions
of temperature, but $\omega$ increases with temperature, so that
$\tau_{crit}$ is larger for higher temperature. For instance,
$\tau_{crit}=2.7$ for $3\times 10^4$K.
In an extreme case of infinite temperature (although unrealistic),
the complete scattering ($\omega=1$) leads to 
$\tau_{crit}= \infty$.
In other words, clouds with any optical depth can be ionized 
due to photon diffusion.

\section{Comparison with Analytic Estimates}

Here, we try to analytically estimate the critical radius based upon
the Str\"{o}mgren approximation.
(A similar estimate is found in Chiba \& Nath 1994.)
By equating the number per unit time of ionizing photons
which enter from the surface to the number per unit time 
of photons which are absorbed in the cloud, 
we have
$R_{HI}=[R^3- 3\pi R^2 I_{\nu_L,0}/h n^2 (1-\chi_{HI})^2$
$\alpha_B(T)] ^{1/3},$
where the ionized regions are assumed to be sharply separated from the
neutral core of radius $R_{HI}$.
Then, the critical radius can be estimated by setting $ R_{HI} =0$:
Resultantly we find
$
R_{crit}=3.5{\rm kpc} \left( M / 10^8 M_\odot \right)^{2/5} 
I_{21}^{-1/5},
$
and the corresponding critical optical depth is $\tau_{crit}=5.3$.
Hence, the dependence of (\ref{rcrit}) upon the cloud mass 
and the UV intensity can be fundamentally understood by this argument. 
But, from a quantitative point of view, this approximation 
obviously leads to overestimation of photoionization effects
as recognized by the critical optical depth.
The overestimation comes from the assumption 
that all photons which enter into the cloud
always cause ionization. In fact, some photons which especially 
have low incident angles do escape from
the gas cloud without causing ionization.
The diffusion process of ionizing photons tends to enhance this effect.
Furthermore, a sharply edged neutral core, which is
the basic assumption in the Str\"{o}mgren approximation, 
is no longer realistic as shown above.

\section{Discussion}

The present results seem of a great significance on considering 
the biasing by photoionization against the formation of low-mass galaxies.
It is shown in previous analyses that if a cloud is assumed to be optically thin,
the photoionization suppresses the collapse of clouds with $M\lsim 10^9 M_\odot$ 
(Umemura \& Ikeuchi 1984) or circular velocities smaller than $30 {\rm ~km~s^{-1}}$
(Thoul \& Weinberg 1996).
In the present analysis, the permeation of UV radiation
is characterized by a different criterion, and 
the critical density has turned out to be almost independent of the mass.
Hence, the evolution of subgalactic clouds would not be determined solely 
by the cloud mass or the circular velocity.
The maximum expansion radius of a top-hat density fluctuation 
is given by 
$R_{max}
%=(4M/3\pi^3 \bar{\rho}_{b,0})^{1/3}(1+z_{max})^{-1}
=10.7{\rm ~kpc}(M/10^8M_\odot)^{1/3}[10/(1+z_{max})]h_{50}^{-2/3}$
in an Einstein-de Sitter universe,
where $z_{max}$ is the maximum expansion epoch, $h_{50}$ is the present
Hubble constant in units of ${\rm 50 km~s^{-1}Mpc^{-1}}$, and
the baryon density parameter is assumed to be 0.05.
Comparing $R_{max}$ with (\ref{rcrit}) and taking into account a possibility
that the UV intensity might be significantly lower at $z>5$, 
we can speculate that the self-shielding could be quite effective 
shortly after the maximum expansion stages. 
Thus, in order to assess the effects of photoionization properly,
we should consider the frequency-dependent 
radiative transfer of diffuse UV photons.

The present results are also relevant to the formation of primordial
hydrogen molecules which provide key physics for the formation of 
the first generation objects (Tegmak et al. 1997).
%When primordial hydrogen molecule formation is considered,
%the first collapsed objects could be considerably small
%($< 10^6 M_\odot$) (Tegmak et al. 1997).
UV radiation will naively suppress the formation of hydrogen 
molecules and thereby cooling (Haiman, Rees, \& Loeb 1997). 
However, when the reionized gas
is self-shielded in the course of evolution, H$^{-}$ or
H$_2^{+}$ ions could form efficiently due to the
residual ionization which may lead to the effective production 
of hydrogen molecules (e.g. Kang \& Shapiro 1992).
Since the formation of primordial hydrogen molecules is a non-equilibrium 
process, the permeation of even a small portion of UV background photons
may play an important role for subsequent molecule formation, and
thereby the evolution of neutral core. 
%This must be also relevant to the formation of low-mass galaxies.

\acknowledgments

We are grateful to T. Nakamoto, and H. Susa for helpful discussion. 
This work was carried out at the Center for Computational Physics 
of University of Tsukuba.
This work was supported in part by the Grants-in Aid of the
Ministry of Education, Science, and Culture, 09874055.

\newpage

\figcaption[Fig.1]{The ionization structure is shown by the neutral
hydrogen fraction $\chi_{HI}$ as a function of radii 
in units of the cloud radius ({\it thick lines}). Here,
$M=10^8M_\odot$ and $I_{21}=1$ are assumed. 
It is seen that the low ionization regions 
rapidly grows due to self-shielding when the radius is smaller than 4.4kpc. 
It is a characteristic 
feature that the boundary between the neutral core and the ionized envelop
is not clear in contrast with the Str\"{o}mgren sphere. This is because
a hard power-law type ($n=-1$) of UV background radiation 
causes appreciable ionization
even in deeper regions by high frequency photons far above the Lyman limit.
Consequently, a small ionization fraction is left in the neutral core.
Comparatively, if an artificial softer spectrum of UV is assumed, a very narrow
transition region appears and the ionization in the neutral core is negligible
({\it thin lines}).
}  

\figcaption[Fig.2]{The central neutral hydrogen fraction is shown
against the cloud radius for a wide variety of cloud mass. $I_{21}=1$
is assumed here.
It is shown that the self-shielding is prominent below a critical
radius (which is defined by the condition that $\chi_{HI}=0.1$ in the
present paper).
}

\figcaption[Fig.3]{The critical radii are plotted as a function of 
UV background intensity.
Different symbols represent different mass of gas clouds.
The straight lines are the fitting formula (see the text).
}

\newpage
\begin{figure}[hbtp]
%\begin{center}
\epsscale{1.0}
\plotone{ion97.fig1}
%\epsfile{file=ion97.fig1,height=20cm}
%\caption{}
%\end{center}
\end{figure}
\vspace{5cm}

\newpage
\begin{figure}[hbtp]
%\begin{center}
\epsfxsize=16cm
\epsfbox{ion97.fig2}
%\epsfile{file=ion97.fig2,height=23cm}
%\caption{}
%\end{center}
\end{figure}

\newpage
\begin{figure}[hbtp]
%\begin{center}
\epsfxsize=16cm
\epsfbox{ion97.fig3}
%\epsfile{file=ion97.fig3,height=23cm}
%\caption{}
%\end{center}
\end{figure}

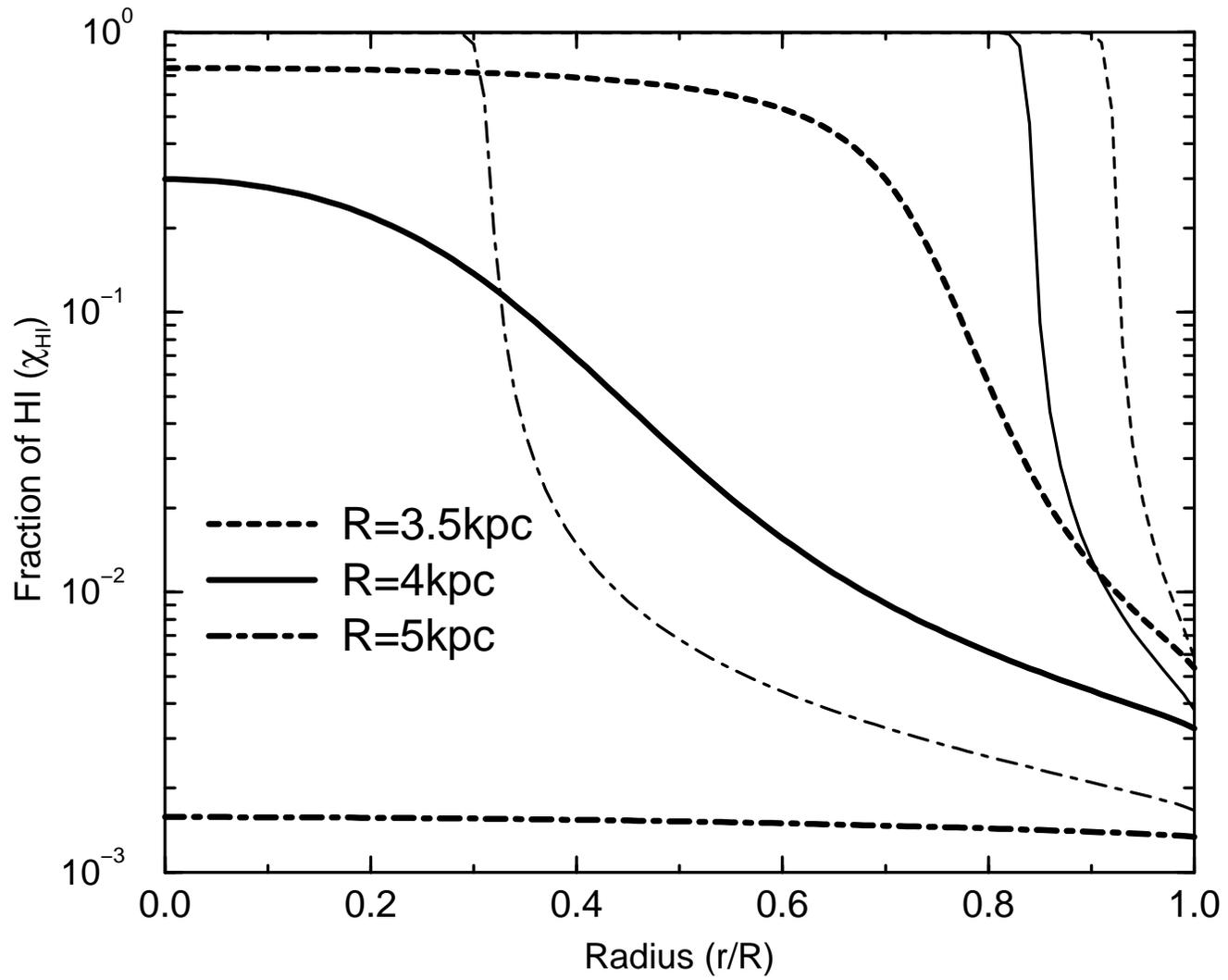
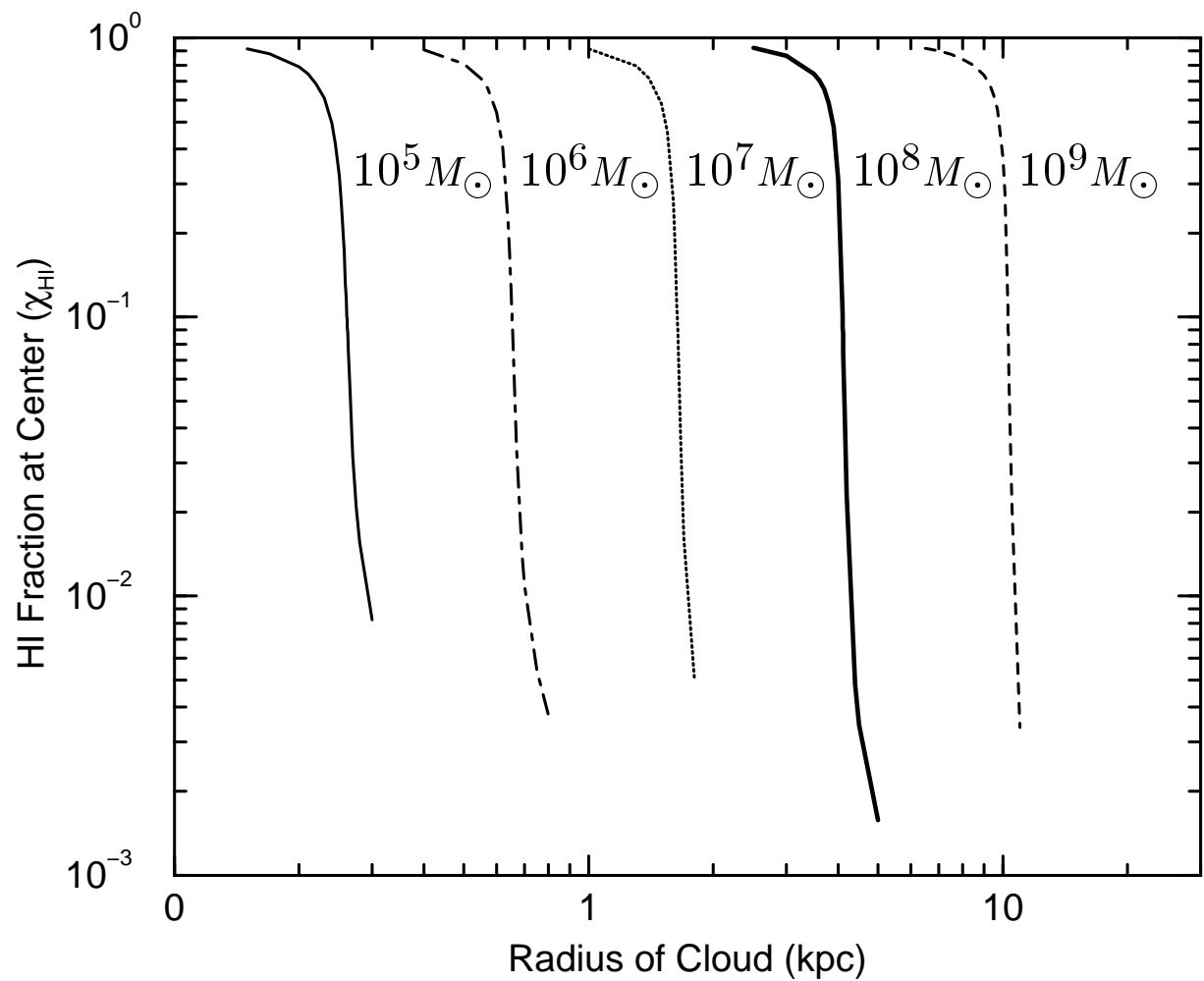
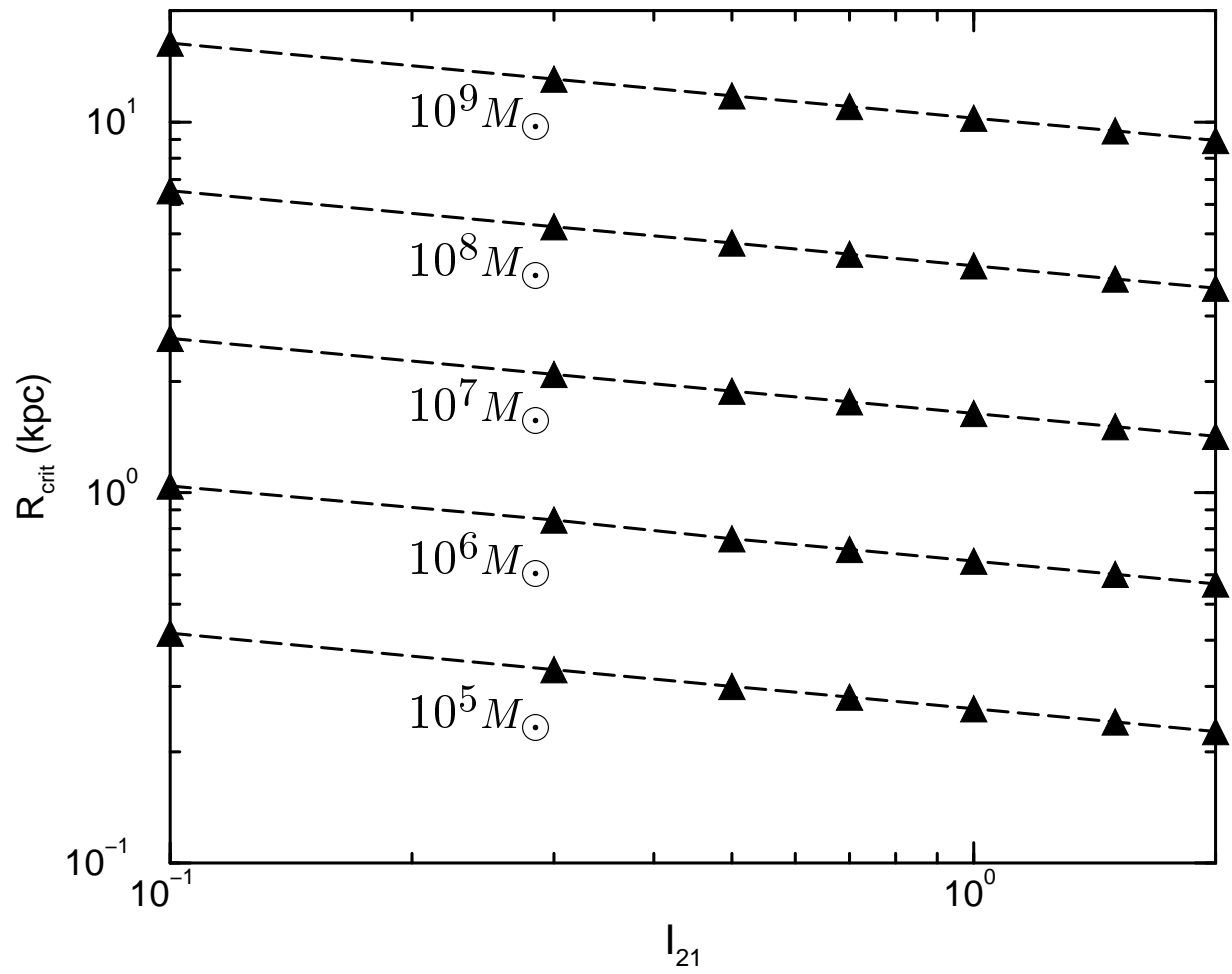
\end{document}